\documentclass[reprint,aps,pra,showpacs,superscriptaddress,longbibliography]{revtex4-1}

\usepackage{graphicx}
\usepackage{dcolumn}
\usepackage{float}
\usepackage{physics}
\usepackage{bm,amsmath}
\usepackage{natbib}
\usepackage{xr}
\usepackage{xcolor}
\usepackage{comment}
\usepackage{array}
\usepackage{float}
\usepackage{lineno}
\usepackage[colorlinks=true, linkcolor=black, citecolor=blue, urlcolor=blue]{hyperref}

\newcommand{\g}{\tilde{g}_\|^0}
\newcommand{\gL}{\tilde{g}_\|^L}
\newcommand{\gR}{\tilde{g}_\|^R}
\newcommand{\gLR}{\tilde{g}_\|^{L(R)}}

\newcommand{\beginsupplement}{
        \setcounter{table}{0}
        \renewcommand{\thetable}{S\arabic{table}}
        \setcounter{figure}{0}
        \renewcommand{\thefigure}{S\arabic{figure}}
        \setcounter{equation}{0}
        \renewcommand{\theequation}{S\arabic{equation}}
        \setcounter{section}{0}
        \renewcommand{\thesection}{S\Roman{section}}
        }

\begin{document}

%%%%%%%%%%%%%%%%%%%%%%%%%%%%%%%%%%
% TITLE
%%%%%%%%%%%%%%%%%%%%%%%%%%%%%%%%%%

\title{Ultra-dispersive resonator readout of a quantum-dot qubit using longitudinal coupling}

%%%%%%%%%%%%%%%%%%%%%%%%%%%%%%%%%%
% AUTHORS 
%%%%%%%%%%%%%%%%%%%%%%%%%%%%%%%%%%

\author{Benjamin Harpt}
\thanks{These two authors contributed equally to this work. Present address: Intel Technology Research, Intel Corporation, Hillsboro, OR 97124, USA.}
\author{J. Corrigan}
\thanks{These two authors contributed equally to this work. Present address: Intel Technology Research, Intel Corporation, Hillsboro, OR 97124, USA.}
\author{Nathan Holman}
\thanks{Present address: HRL Laboratories, LLC., 3011 Malibu Canyon Road, Malibu, CA 90265, USA.}
\author{Piotr Marciniec}
\affiliation{Department of Physics, University of Wisconsin, Madison, WI 53706, USA}

\author{D. Rosenberg}
\author{D. Yost}
\author{R. Das}
\affiliation{MIT Lincoln Laboratory, 244 Wood Street, Lexington, MA 02421, USA}

\author{Rusko Ruskov}
\author{Charles Tahan}
\affiliation{Laboratory for Physical Sciences, 8050 Greenmead Dr., College Park, MD 20740, USA}

\author{William D. Oliver}
\affiliation{MIT Lincoln Laboratory, 244 Wood Street, Lexington, MA 02421, USA}

\author{R. McDermott}
\author{Mark Friesen}
\author{M. A. Eriksson}
\thanks{\href{mailto:maeriksson@wisc.edu}{maeriksson@wisc.edu}}
\affiliation{Department of Physics, University of Wisconsin, Madison, WI 53706, USA}

\date{\today}

%%%%%%%%%%%%%%%%%%%%%%%%%%%%%%%%%%
% ABSTRACT 
%%%%%%%%%%%%%%%%%%%%%%%%%%%%%%%%%%

\begin{abstract}
    We perform readout of a quantum-dot hybrid qubit coupled to a superconducting resonator through a parametric, longitudinal interaction mechanism.
    Our experiments are performed with the qubit and resonator frequencies detuned by $\sim$10 GHz, demonstrating that longitudinal coupling can facilitate semiconductor qubit operation in the `ultra-dispersive' regime of circuit quantum electrodynamics.
\end{abstract}

\maketitle

%%%%%%%%%%%%%%%%%%%%%%%%%%%%%%%%%%
% INTRODUCTION
%%%%%%%%%%%%%%%%%%%%%%%%%%%%%%%%%%

Coupling semiconductor qubits to superconducting resonators using circuit quantum electrodynamics (cQED) techniques is an important element in proposed quantum computing architectures \cite{Childress.2004.10.1103/physreva.69.042302, Vandersypen.2017.10.1038/s41534-017-0038-y}.
Transverse coupling schemes based on qubit-resonator photon exchange are conventionally studied for this purpose.
However, the transverse interaction becomes weak when the qubit and resonator frequencies are far from resonance, which could constrain operation protocols for large-scale processors.
Longitudinal interactions offer an intriguing alternative for quantum coupling, as they do not require frequency resonance \cite{Kerman.2013}.
These couplings have attracted growing attention, and although they have recently been observed in semiconductor-qubit systems \cite{Bottcher.2022, Corrigan.2023.PhysRevApplied.20.064005, champain2024parametriclongitudinalcouplingsemiconductor}, they have not yet been utilized to extend device functionality.
Here, as an initial use case for longitudinal coupling in quantum-dot qubits, we demonstrate excited-state resonator readout and spectroscopy at qubit frequencies far beyond the typical cQED operating range.

%%%%%%%%%%%%%%%%%%%%%%%%%%%%%%%%%%
% DEVICE & MEASUREMENT SETUP
%%%%%%%%%%%%%%%%%%%%%%%%%%%%%%%%%%

\begin{figure*}[t]
    \centering
    \includegraphics[width=1\linewidth, trim=1.4cm 2.1cm 1.5cm 1.7cm, clip]{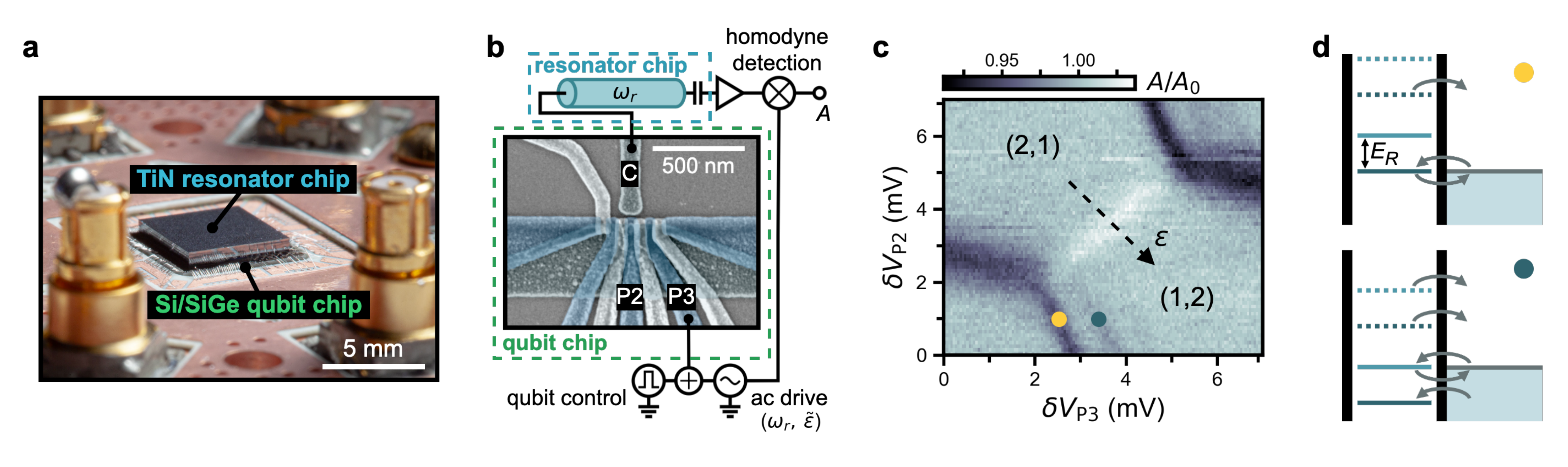}
    \caption{
        Pulsed-gate spectroscopy of a 3D DQD-resonator device. 
        \textbf{(a)} Photo of the packaged sample, comprising a Si/SiGe quantum-dot qubit device and TiN microwave resonator integrated in a flip-chip architecture.
        \textbf{(b)} Simplified circuit schematic for experiments. A false-color, scanning electron micrograph of a quantum-dot device nominally identical to the one used in experiments is shown. A DQD is formed under gates P2 and P3 and capacitively coupled  to the off-chip resonator via gate C. Qubit control pulses and an ac drive at frequency $\omega_r$ are applied to P3 (dc voltage sources are not shown). The resonator transmission amplitude $A$ is measured using homodyne detection.
        \textbf{(c)} Normalized transmission amplitude $A / A_0$ measured as a function of dc plunger voltages $V_\text{P2}$ and $V_\text{P3}$ near the (2,1)-(1,2) charge transition. A pulsed gate voltage is applied for spectroscopy of the right dot. Reservoir tunneling resonances with the lowest two energy states are marked with yellow and teal circles.
        \textbf{(d)} Energy level configurations corresponding to the equipotentials marked in (c). The dot's energy levels are cycled between lower (solid) and upper (dashed) positions by the alternating-voltage waveform. Arrows indicate electron tunneling pathways between the dot and reservoir.
        }
    \label{fig:device}
\end{figure*}

Experiments are performed using a Si/SiGe quantum-dot qubit device coupled to an off-chip, TiN microwave resonator.
The qubit and resonator are fabricated on separate dies and vertically integrated in a flip-chip architecture, shown in Fig.~\ref{fig:device}(a) \cite{Holman.2021.10.1038/s41534-021-00469-0, Corrigan.2023.PhysRevApplied.20.064005}.
Figure~\ref{fig:device}(b) shows a scanning electron micrograph of a nominally-identical device with a simplified schematic of the measurement circuit.
A double quantum dot (DQD) formed under plunger gates P2 and P3 is coupled to the resonator field through capacitance with gate C; the estimated charge-photon coupling rate is $g_c / 2\pi \approx 3.2$~MHz.

During experiments, a continuous ac drive is applied to P3 at the resonator fundamental-mode frequency of $\omega_r / 2 \pi \approx 1.304$~GHz. 
This drive modulates the DQD energy detuning adiabatically with amplitude $\tilde{\varepsilon}$, activating a dynamic longitudinal coupling with the resonator photons \cite{Kerman.2013, Ruskov.2019.10.1103/physrevb.99.245306, Bottcher.2022, Corrigan.2023.PhysRevApplied.20.064005, champain2024parametriclongitudinalcouplingsemiconductor}.
The effective interaction Hamiltonian is
\begin{equation}
    \tilde{H}_\| = \hbar \, \g \cos{\left( \omega_r t \right)} \, \hat{\sigma}_z \left( \hat{a} + \hat{a}^\dagger \right),
    \label{eq:H_int}
\end{equation}
where $\g$ denotes the coupling strength, $\hat{\sigma}_z$ is the Pauli $z$ operator in the qubit energy basis, and $\hat{a}$ and $\hat{a}^\dagger$ are photon annihilation and creation operators.
The coupling $\g$ gives rise to a qubit-state-dependent resonator drive, which can be measured through changes in its stationary field amplitude.
Figure~\ref{fig:device}(c) shows the normalized resonator transmission amplitude $A / A_0$ measured near the (2,1)-(1,2) interdot charge transition.
The axis for DQD detuning $\varepsilon$ is indicated, where the total detuning, $\varepsilon(t) = \varepsilon_0 + \tilde{\varepsilon} \cos{(\omega_r t)}$, includes a dc component $\varepsilon_0$.
The normalization constant $A_0$ is defined by the transmission level at $\varepsilon_0 \gg 0$.
A boost in $A/A_0$ measured around $\varepsilon_0 = 0$ is a telltale indicator of $\g$ coupling the DQD to the resonator for this measurement configuration, as shown in Ref.~\cite{Corrigan.2023.PhysRevApplied.20.064005}.

%%%%%%%%%%%%%%%%%%%%%%%%%%%%%%%%%%
% ULTRA-DISPERSIVE REGIME
%%%%%%%%%%%%%%%%%%%%%%%%%%%%%%%%%%

In the following experiments, we operate our device as a quantum-dot hybrid qubit (QDHQ) \cite{Shi.2012.10.1103/physrevlett.108.140503, Kim.2014.10.1038/nature13407, cao.2016.tunable, Thorgrimsson.2017.Extending, jang.2021.single-shot, park2024singleshotlatchedreadoutquantum}. 
The qubit excitation frequency at $\varepsilon_0 = 0$ in Fig.~\ref{fig:device}(c) is $\omega_q / 2\pi \approx 2 \Delta_1 / h = 14.6 \pm 1.2$~GHz, where the interdot tunnel coupling $\Delta_1$ is extracted through fits to the resonator transmission peak (see the Supplementary Information \cite{supplement} for further details).
At large $|\varepsilon_0|$, $\omega_q$ is set by the left- or right-dot singlet-triplet splitting, $E_L$ or $E_R$.
We measure $E_R$ using time-averaged pulsed-gate spectroscopy \cite{Elzerman.2004.10.1063/1.1757023}, with an alternating-voltage waveform applied to P3.
Reservoir tunneling resonances for the ground and first-excited states are marked in Fig.~\ref{fig:device}(c) with yellow and teal circles, respectively.
These resonances occur along equipotentials where a dot energy level aligns with the reservoir Fermi energy, as depicted in Fig.~\ref{fig:device}(d).
From the voltage splitting between resonances, we estimate $E_R / h = 34.8 \pm 1.2$~GHz~$\approx \omega_q / 2 \pi$ for $\varepsilon_0 \gg 0$ \cite{supplement}.
We note that the magnitude of $\tilde{\varepsilon}$ is kept much smaller than $\hbar \omega_q$ and all tunnel couplings for our measurements, so the ac drive cannot excite the qubit. 

The above measurements demonstrate operation in the `ultra-dispersive' coupling regime, which we define by the condition $g_c \ll \omega_r \ll \omega_q$.
Dynamic longitudinal coupling is an important mechanism for working in this regime; through parametric driving, its coupling strength, $\g \propto (\partial^2 E_q / \partial \varepsilon^2) \, \tilde{\varepsilon}$, where $(\partial^2 E_q / \partial \varepsilon^2)$ is the qubit's energy curvature, can be made significantly larger than the effective transverse coupling, which is considerably weakened at $\omega_q \gg \omega_r$ \cite{Corrigan.2023.PhysRevApplied.20.064005, champain2024parametriclongitudinalcouplingsemiconductor, Ruskov.2019.10.1103/physrevb.99.245306, Guo.2023}.
For this work, we estimate longitudinal coupling to be the dominant interaction by an order of magnitude \cite{supplement}, and in the experiments below, we show how this coupling can be used for QDHQ state readout in the ultra-dispersive regime.

%%%%%%%%%%%%%%%%%%%%%%%%%%%%%%%%%%
% READOUT EXPERIMENTS
%%%%%%%%%%%%%%%%%%%%%%%%%%%%%%%%%%

\begin{figure*}[t]
    \centering
    \includegraphics[width=0.85\linewidth, trim=0cm 1.7cm 0cm 1.8cm, clip]{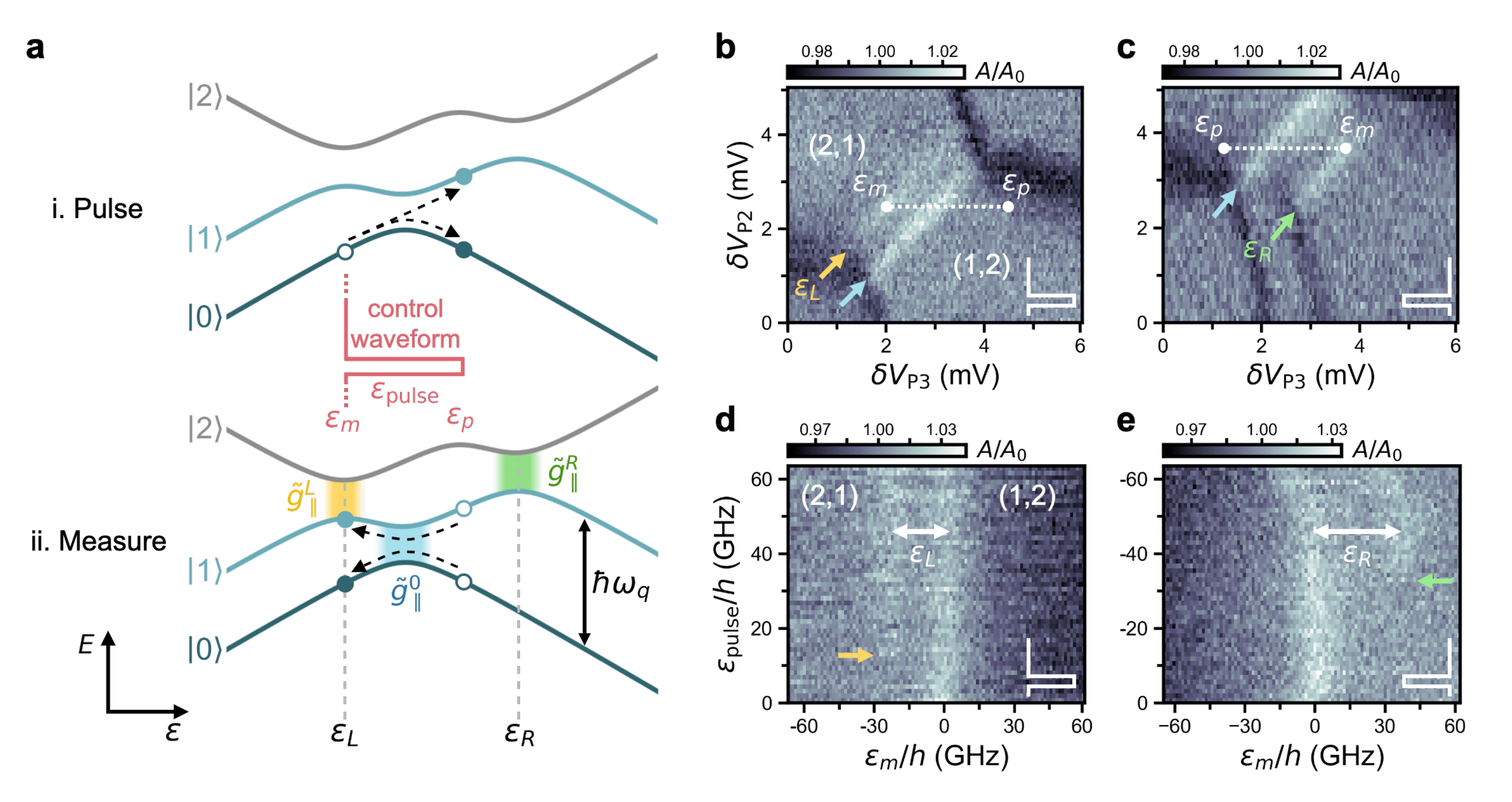}
    \caption{
        QDHQ readout and spectroscopy using longitudinal coupling.
        \textbf{(a)} QDHQ energy level diagram and qubit control scheme. 
        The pulsed voltage waveform is applied to P3. 
        Dashed lines illustrate qubit dynamics and state transitions.
        \textbf{(b, c)} $A / A_0$ measured as a function of dc voltages $V_\text{P2}$ and $V_\text{P3}$ with positive-amplitude [panel (b)] and negative-amplitude [panel (c)] control pulses applied simultaneously. 
        Dashed white lines indicate the scale of detuning pulses. Signal boosts marked with arrows are from longitudinal coupling at the corresponding-color (blue, yellow, and green) anticrossings in (a).
        \textbf{(d, e)} $A / A_0$ as a function of measurement detuning $\varepsilon_m$ and pulse amplitude $\varepsilon_\text{pulse}$. 
        Arrows highlight $\ket{1}$-state occupation thresholds. 
        The $\ket{1}$-$\ket{2}$ energy anticrossing detunings $\varepsilon_L$ and $\varepsilon_R$ can be estimated from the distance between transmission peaks.
    }
    \label{fig:readout}
\end{figure*}

Figure~\ref{fig:readout}(a) illustrates the QDHQ energy levels and a two-step qubit control sequence with fast voltage pulses applied to P3.
The logical states $\ket{0}$ and $\ket{1}$ and leakage state $\ket{2}$ are defined as energy eigenstates for the lowest three levels.
In the first control step [panel (i)], $\varepsilon_0$ is pulsed from $\varepsilon_m$ to $\varepsilon_p$ for an interval of $t_\text{pulse} = 2$~ns.
The detuning amplitude of the pulse is given by $\varepsilon_\text{pulse} = \varepsilon_p - \varepsilon_m$.
As the pulse shuttles the qubit through the $\ket{0}$-$\ket{1}$ energy anticrossing at $\varepsilon_0 = 0$, fractional excited-state occupation is generated through Landau-Zener transitions \cite{Petta.2010.10.1126/science.1183628, Shi.2013.10.1038/ncomms4020}.
In the second control step [panel (ii)], $\varepsilon_0$ is pulsed back to $\varepsilon_m$ and maintained there for a measurement time of $t_\text{meas} = 48$~ns.
Qubit-resonator coupling during this interval generates the transmission signal used for readout.
We note that no coherent oscillations are observed in our measurements, likely because the manipulation scheme frequently involves idling in the detuning range $\Delta_1 < |\varepsilon_0| < E_{L(R)}$, where the qubit is highly susceptible to charge noise \cite{Kim.2014.10.1038/nature13407, Thorgrimsson.2017.Extending}.
The control protocol can be easily modified to avoid these detuning zones in future experiments. 

The above pulse sequence is cycled while simultaneously sweeping the dc voltages $V_\text{P2}$ and $V_\text{P3}$ near the (2,1)-(1,2) transition.
Figures~\ref{fig:readout}(b) and \ref{fig:readout}(c) show time-averaged $A / A_0$ measurements with opposite-amplitude pulses applied, $\varepsilon_\text{pulse} / h \approx 63$~GHz and $-63$~GHz, respectively; these amplitudes are indicated by dashed white lines.
In both plots, a peak in $A / A_0$ is visible at $\varepsilon_m = 0$ (blue arrows), where the adiabatic qubit modulation near the heightened energy curvature of the $\ket{0}$-$\ket{1}$ anticrossing enlarges $\g$ \cite{Corrigan.2023.PhysRevApplied.20.064005, Ruskov.2019.10.1103/physrevb.99.245306}.
Additional boosts in $A / A_0$ at $|\varepsilon_m| \neq 0$ (yellow and green arrows) are the result of resonator coupling through occupation of the first excited state.
These features are caused by qubit modulation near a $\ket{1}$-$\ket{2}$ anticrossing at $\varepsilon_m = \varepsilon_L$ [Fig.~\ref{fig:readout}(b)] or $\varepsilon_m = \varepsilon_R$ [Fig.~\ref{fig:readout}(c)], activating an excited-state coupling channel, which we label $\gL$ and $\gR$, respectively.
DQD-resonator coupling involving excited silicon valley states has been observed previously, but was mediated through the transverse interaction  \cite{Mi.2017.10.1103/physrevlett.119.176803}; the $A / A_0$ peak in our measurements is the key signature of longitudinal coupling.

A crucial feature of the couplings $\gL$ and $\gR$ is that they are only active when the logical state $\ket{1}$ is populated. 
As a consequence, the resonator transmission signal observed at $\varepsilon_L$ or $\varepsilon_R$ reflects the qubit's time-averaged $\ket{1}$ occupation.
Figures~\ref{fig:readout}(d) and \ref{fig:readout}(e) show experiments demonstrating that this signal can be used to detect the qubit state.
The $A/A_0$ peaks are measured as a function of $\varepsilon_m$ and $\varepsilon_\text{pulse}$.
For small pulse amplitudes ($|\varepsilon_\text{pulse}| \ll |\varepsilon_{L(R)}|$), the qubit's prepared state is $\ket{0}$, and no signal boost occurs at $\varepsilon_{L(R)}$.
When the control pulse becomes large enough to generate excitation into $\ket{1}$, the prepared state is no longer $\ket{0}$, and the signal at $\varepsilon_{L(R)}$ increases via the excited-state resonator coupling $\gLR$.
In our measurements, the $\varepsilon_L$ peak in Fig.~\ref{fig:readout}(d) becomes visible at a threshold amplitude of $\varepsilon_\text{pulse} / h \approx 13$~GHz (yellow arrow), and the $\varepsilon_R$ peak in Fig.~\ref{fig:readout}(e) appears at $\varepsilon_\text{pulse} / h \approx -33$~GHz (green arrow).
The dependence of the signal at $\varepsilon_L$ or $\varepsilon_R$ on the qubit's prepared state implies that QDHQ readout can be performed at these working points using longitudinal coupling.

The data in Fig.~\ref{fig:readout} contain spectroscopic information about the qubit's energy parameters.
Using these measurements and additional numerical analysis methods outlined in \cite{supplement}, we are able to reconstruct the full QDHQ Hamiltonian and compute the qubit frequency.
We estimate $\omega_q / 2 \pi = 20.6 \pm 0.6$~GHz and $32.4 \pm 2.9$~GHz at the $\varepsilon_L$ and $\varepsilon_R$ readout operating points.
This affirms the ultra-dispersive nature of our experiments, since qubit-resonator detuning in this regime, $\Delta = |\omega_q - \omega_r| \sim 2 \pi \times 10$~GHz, is one to three orders of magnitude larger than typically used for dispersive-cQED experiments \cite{Wallraff.2005.PhysRevLett.95.060501, Majer.2007.Coupling, Mi.2018.coherent, Dijkema.2023.Two-qubit}.

%%%%%%%%%%%%%%%%%%%%%%%%%%%%%%%%%%
% DISCUSSION
%%%%%%%%%%%%%%%%%%%%%%%%%%%%%%%%%%

The experiments above outline a method for ultra-dispersive QDHQ readout based on longitudinal coupling. 
This approach differs from charge-based QDHQ readout \cite{Kim.2014.10.1038/nature13407, cao.2016.tunable, Thorgrimsson.2017.Extending, jang.2021.single-shot, park2024singleshotlatchedreadoutquantum}, as well as from dispersive readout using transverse coupling, which has been demonstrated with other spin qubit encodings \cite{Mi.2018.coherent, Dijkema.2023.Two-qubit, croot.2020.flopping-mode}. 
In comparison, our method offers several potential benefits for qubit operation, which can be explored in future experiments.
First, it implements, in principle, a quantum-nondemolition measurement, as $\tilde{H}_\|$ [Eq.~(\ref{eq:H_int})] commutes with the qubit Hamiltonian \cite{Ruskov.2019.10.1103/physrevb.99.245306, ruskov.2024.longitudinal}.
Second, it requires no qubit-resonator photon exchange, minimizing pathways for Purcell decay \cite{Houck.2008.10.1103/physrevlett.101.080502} and measurement-induced dephasing \cite{Slitcher.2012.10.1103/physrevlett.109.153601} intrinsic to transverse interactions \cite{Kerman.2013}.
As a result, it is possible to operate at higher photon populations without perturbing the qubit, facilitating fast and high-fidelity readout \cite{didier.2015.fast, ikonen.2019.qubit}.
We also expect the probability of state transitions due to the ac qubit drive to be low, provided that it is weak and adiabatic.
Third, our readout is performed at DQD detuning points where qubit relaxation involves a slow spin-flip mechanism, which may provide a $T_1$ advantage compared to reading out at the $\varepsilon_0 = 0$ measurement point, where no spin flip is required.
We emphasize that slow qubit relaxation could be helpful for improving measurement fidelity and achieving single-shot readout in near-term experiments, where relatively weak coupling may necessitate longer integration times.
Finally, our technique works in the ultra-dispersive regime, where transverse coupling is suppressed.
Our work demonstrates that qubits can be measured through longitudinal interactions while far-detuned from the readout resonator, allowing space for a wide range of qubit frequencies.
Longitudinal coupling, therefore, is a promising tool for mitigating engineering bottlenecks and scaling quantum processors.

%%%%%%%%%%%%%%%%%%%%%%%%%%%%%%%%%%
% ACKNOWLEDGEMENTS
%%%%%%%%%%%%%%%%%%%%%%%%%%%%%%%%%%

\section*{Acknowledgements}

The authors thank HRL Laboratories for support and L.~F.~Edge for providing the Si/Si$_{1-x}$Ge$_x$ heterostructure used in this work. 
This research was sponsored in part by the Army Research Office under Awards No.~W911NF-17-1-0274 and No.~W911NF-23-1-0110.
J.C.~acknowledges support from the U.S.~National Science Foundation Graduate Research Fellowship Program under Grant No.~DGE-1747503 and the Graduate School and the Office of the Vice Chancellor for Research and Graduate Education at the University of Wisconsin-Madison with funding from the Wisconsin Alumni Research Foundation. 
The authors acknowledge the use of facilities supported by the National Science Foundation through the University of Wisconsin-Madison Materials Research Science and Engineering Center (Grant No.~DMR-2309000) and the Major Research Instrumentation (MRI) program (Grant No.~DMR-1625348). 
This material is based in part upon work supported by the Under Secretary of Defense for Research and Engineering under Air Force Contract No.~FA8702-15-D-0001. 
Any opinions, findings, conclusions or recommendations expressed in this material are those of the author(s) and do not necessarily reflect the views of the Under Secretary of Defense for Research and Engineering and are not necessarily endorsed by nor should they be interpreted as representing the official policies, either expressed or implied, of the Army Research Office or the U.S.~Government. 
The U.S.~Government is authorized to reproduce and distribute reprints for U.S.~Government purposes notwithstanding any copyright notation herein.

%%%%%%%%%%%%%%%%%%%%%%%%%%%%%%%%%%
% DATA AVAILABILITY
%%%%%%%%%%%%%%%%%%%%%%%%%%%%%%%%%%

\section*{Data availability}

The data that support the findings of this study are openly available in a \href{https://doi.org/10.5281/zenodo.14187719}{Zenodo repository} \cite{zenodo}.

\section*{Code availability}

Analysis code for the spectroscopy measurements in this study is available in a \href{https://doi.org/10.5281/zenodo.14187719}{Zenodo repository} \cite{zenodo}.

%%%%%%%%%%%%%%%%%%%%%%%%%%%%%%%%%%
% COMPETING INTERESTS
%%%%%%%%%%%%%%%%%%%%%%%%%%%%%%%%%%

\section*{Competing interests}

The authors declare no financial or non-financial competing interests.

%%%%%%%%%%%%%%%%%%%%%%%%%%%%%%%%%%
% AUTHOR CONTRIBUTIONS
%%%%%%%%%%%%%%%%%%%%%%%%%%%%%%%%%%

\section*{Author contributions}

N.H., D.R., D.Y., R.D., and W.D.O.~fabricated the flip-chip sample used in experiments.
B.H.~and J.C.~performed all measurements and data analysis, with N.H., P.M., R.M., and M.A.E.~providing assistance.
R.R., C.T., and M.F.~developed theoretical models used for data analysis.
B.H., J.C., M.F., and M.A.E.~wrote the manuscript, with input from all authors.

%%%%%%%%%%%%%%%%%%%%%%%%%%%%%%%%%%
% REFERENCES
%%%%%%%%%%%%%%%%%%%%%%%%%%%%%%%%%%

%

%%%%%%%%%%%%%%%%%%%%%%%%%%%%%%%%%%
% SUPPLEMENTARY INFO
%%%%%%%%%%%%%%%%%%%%%%%%%%%%%%%%%%

\onecolumngrid
\clearpage

\beginsupplement

\section*{Supplementary Information}

These supplemental materials provide additional data and details about the methods used in this work.

%%%%%%%%%%%%%%%%%%%%%%%%%%%%%%%%%%
% QUBIT HAMILTONIAN
%%%%%%%%%%%%%%%%%%%%%%%%%%%%%%%%%%

\section{Qubit Hamiltonian \& Logical States}

\begin{figure}[b]
    \centering
    \includegraphics[width=0.65\linewidth, trim=0cm 2.6cm 0cm 2.6cm, clip]{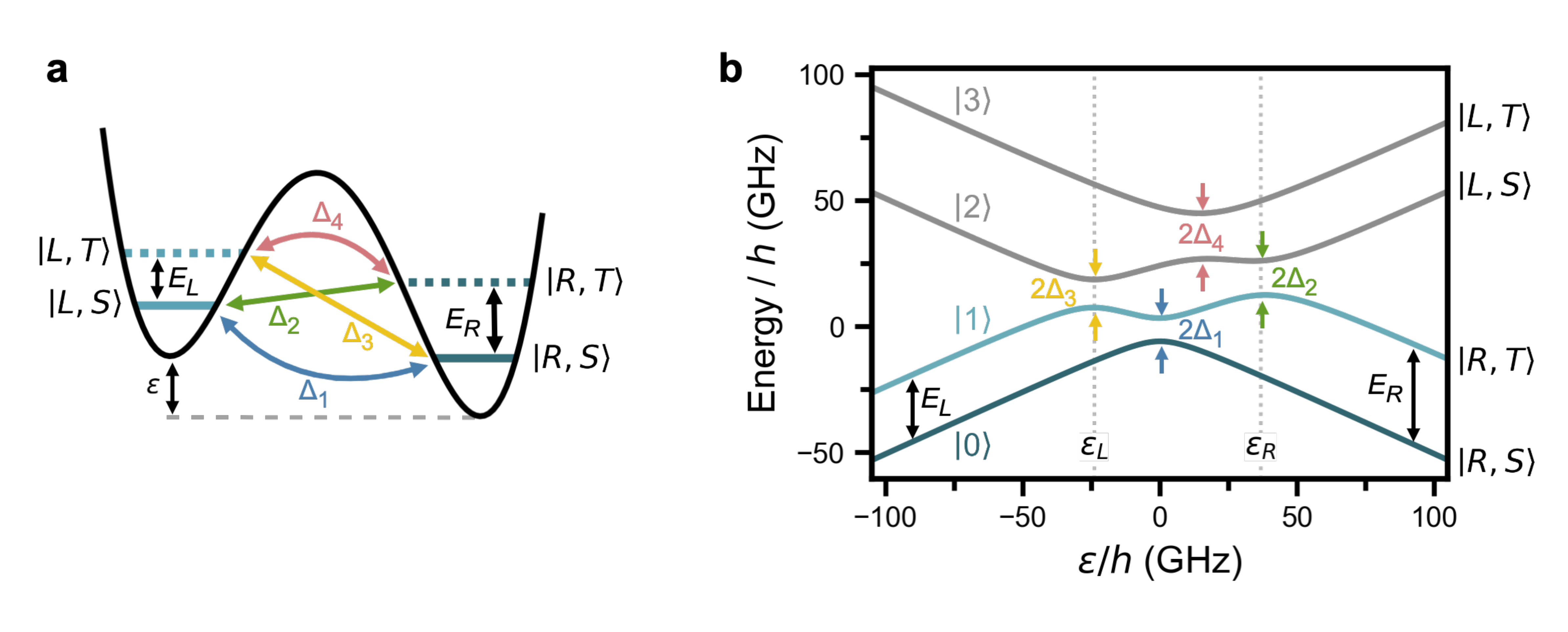}
    \caption{
        QDHQ states and energy levels.
        \textbf{(a)} Electrostatic potential for a three-electron DQD.
        Quantum-dot energy levels are labeled corresponding to the basis states of Eq.~(\ref{eq:QDHQstates}); the third electron's level occupation determines the state of the system in this basis.
        $E_L$ and $E_R$ denote the singlet-triplet splitting for each dot, and $\Delta_1$ through $\Delta_4$ are interdot tunnel couplings.
        \textbf{(b)} Energy levels of the system in (a) as a function of $\varepsilon$.
        Qubit logical-state levels are colored teal.
        The energy eigenstates $\ket{0}$ through $\ket{3}$ are asymptotically equal to the basis states in Eq.~(\ref{eq:QDHQstates}) at large $|\varepsilon|$.
        Detuning values of the $\ket{1}$-$\ket{2}$ anticrossings are labeled $\varepsilon_L$ and $\varepsilon_R$.
    }
    \label{fig:QDHQ}
\end{figure}

The QDHQ combines the charge and spin degrees of freedom of a three-electron DQD depicted in Fig.~\ref{fig:QDHQ}(a). 
The DQD is operated with two electrons in one dot and a single electron in the other dot.
In the absence of tunnel coupling, the eigenstates of the four-level system pictured can be defined in the charge-spin basis:
\begin{equation}
    \begin{alignedat}{2}
        & \ket{L,S} \equiv \ket{S}\ket{\downarrow}; \qquad
        && \ket{L,T} \equiv \sqrt{\frac{1}{3}} \ket{T_0} \ket{\downarrow} - \sqrt{\frac{2}{3}} \ket{T_-} \ket{\uparrow}; \\
        & \ket{R,S} \equiv \ket{\downarrow}\ket{S}; \qquad
        && \ket{R,T} \equiv \sqrt{\frac{1}{3}} \ket{\downarrow} \ket{T_0} - \sqrt{\frac{2}{3}} \ket{\uparrow} \ket{T_-}.
    \end{alignedat}
    \label{eq:QDHQstates}
\end{equation}
In these definitions, $\ket{S} \equiv (\ket{\uparrow \downarrow} - \ket{\downarrow \uparrow}) / \sqrt{2}$ is the spin-singlet state, while $\ket{T_0} \equiv (\ket{\uparrow \downarrow} + \ket{\downarrow \uparrow}) / \sqrt{2}$ and $\ket{T_-} \equiv \ket{\downarrow \downarrow}$ are spin-triplet states of the doubly-occupied dot; $\ket{\uparrow}$ and $\ket{\downarrow}$ denote spin-up and spin-down electrons in the singly-occupied dot.

The bare (i.e., undressed) DQD Hamiltonian in the charge-spin basis ordered \{$\ket{L,S}, \ket{L,T}, \ket{R,S}, \ket{R,T}$\} is
\begin{equation}
    \renewcommand{\arraystretch}{1.2}
    H_q = 
    \begin{pmatrix}
        \varepsilon / 2 & 0 & \Delta_1 & -\Delta_2 \\
        0 & \varepsilon / 2 + E_L & -\Delta_3 & \Delta_4 \\
        \Delta_1 & -\Delta_3 & -\varepsilon / 2 & 0 \\
        -\Delta_2 & \Delta_4 & 0 & -\varepsilon / 2 + E_R 
    \end{pmatrix},
    \label{eq:H_qdhq}
\end{equation}
where $E_L$ and $E_R$ are the singlet-triplet energy level splittings in the left and right dot, respectively. 
There are four interdot tunnel coupling parameters, denoted $\Delta_1$ through $\Delta_4$, which lead to hybridization of the eigenstates in Eq.~(\ref{eq:QDHQstates}).
Figure~\ref{fig:QDHQ}(b) plots the energy levels of $H_q$ as a function of $\varepsilon$ for our device, with parameter values derived in the following sections. 
For this work, we define the qubit logical states $\ket{0}$ and $\ket{1}$ as the two lowest-energy eigenstates of $H_q$ [teal-colored energy levels in Fig.~\ref{fig:QDHQ}(b)]. 
At large DQD detuning, these logical states are asymptotically equal to the basis states in Eq.~(\ref{eq:QDHQstates}).

%%%%%%%%%%%%%%%%%%%%%%%%%%%%%%%%%%
% ANALYSIS METHODS
%%%%%%%%%%%%%%%%%%%%%%%%%%%%%%%%%%

\section{Qubit Spectroscopy Analysis Methods}

The following sections describe analysis procedures used for estimating the QDHQ energy parameters. 
Before plotting or analysis, all data are normalized by the constant $A_0$, which denotes the transmission amplitude measured with the DQD in Coulomb blockade.
$A_0$ is determined independently for each dataset. The estimated values and uncertainties of all parameters are summarized in Table~\ref{tab:params} below.

%%%%%%%%%%%%%%%%%%%%%%%%%%%%%%%%%%

\subsection{Quantum-dot gate lever arms}

The device gate lever arms convert changes in gate voltage to calibrated movement of the quantum-dot chemical potentials, and are used throughout the analyses below. 
We measure gate-to-dot lever arms from the thermal broadening of reservoir tunneling transition lines, as detailed in Ref.~\cite{Corrigan.2023.PhysRevApplied.20.064005}, obtaining $\alpha_\text{P2}^L \equiv |\partial\mu_L / \partial V_\text{P2}| = 0.140 \pm 0.005$~eV/V and $\alpha_\text{P3}^R \equiv |\partial\mu_R / \partial V_\text{P3}| = 0.149 \pm 0.003$~eV/V, where $\mu_L$ and $\mu_R$ are the chemical potentials of the left and right dot. The DQD detuning lever arms for gates P3 and C are calculated using the slopes of reservoir transitions in two-dimensional voltage sweeps:
\begin{equation}
    \alpha_\text{P3}^\varepsilon \equiv \frac{\partial\varepsilon_0}{\partial V_\text{P3}} = \alpha_\text{P3}^R - \alpha_\text{P2}^L \left| \frac{\delta V_\text{P2}}{\delta V_\text{P3}} \right|_L = 0.105 \pm 0.005~\text{eV/V};
    \label{eq:alphaP3}
\end{equation}
\begin{equation}
    \alpha_\text{C}^\varepsilon \equiv \left| \frac{\partial\varepsilon_0}{\partial V_\text{C}} \right| = \alpha_\text{P2}^L \left| \frac{\delta V_\text{P2}}{\delta V_\text{C}} \right|_L - \alpha_\text{P3}^R \left| \frac{\delta V_\text{P3}}{\delta V_\text{C}} \right|_R = 0.023 \pm 0.004~\text{eV/V}.
    \label{eq:alphaC}
\end{equation}
In these equations, the quantity $|\delta V_i / \delta V_j|_{L(R)}$ represents the slope of a left-dot (right-dot) reservoir transition in $(V_j,V_i)$ parameter space. 
The values of all lever arms are found to remain roughly constant between the main-text device tunings of Fig.~\ref{fig:device}(c) and Fig.~\ref{fig:readout}.

%%%%%%%%%%%%%%%%%%%%%%%%%%%%%%%%%%

\subsection{\texorpdfstring{Interdot tunnel couplings \& $\bm{|1\rangle}$-$\bm{|2\rangle}$ anticrossing detunings: $\bm{\Delta_1, \, \Delta_2, \, \Delta_3, \, \varepsilon_L, \, \varepsilon_R}$}{Interdot tunnel couplings and |1>-|2> anticrossing detunings: Δ1, Δ2, Δ3, εL, εR}}

\begin{figure}[t]
    \centering
    \includegraphics[width=0.55\linewidth]{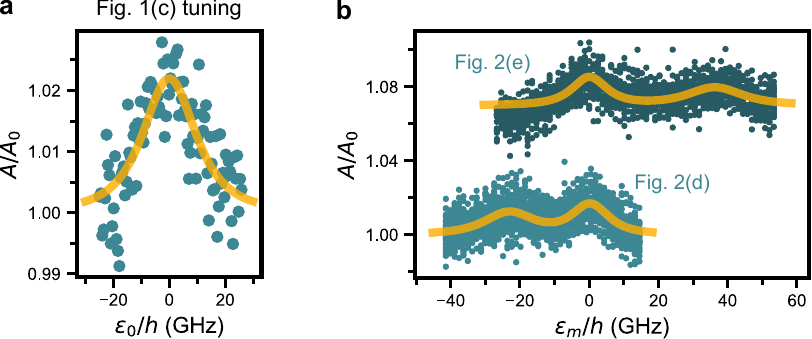}
    \caption{
        Curve-fitting analysis of resonator transmission peaks.
        \textbf{(a)} $A/A_0$ near the $\ket{0}$-$\ket{1}$ anticrossing, measured at the same device tuning as Fig.~\ref{fig:device}(c).
        The data are fit using Eq.~(\ref{eq:Anorm}) (yellow curve).
        \textbf{(b)} Combined traces from Figs.~\ref{fig:readout}(d) (light teal points) and \ref{fig:readout}(e) (dark teal points; offset for clarity).
        The data are fit using Eq.~(\ref{eq:Anormmulticoupling}) (yellow curves).
    }
    \label{fig:tunnelcoupling}
\end{figure}

The interdot tunnel coupling parameters can be extracted by fitting the resonator transmission peaks generated by longitudinal coupling at the QDHQ energy anticrossings.
Figure~\ref{fig:tunnelcoupling}(a) shows an $A / A_0$ peak acquired near the $\ket{0}$-$\ket{1}$ anticrossing at the same device tuning used for Fig.~\ref{fig:device}(c) of the main text.
We do not apply any control pulses during this measurement; consequently, the DQD remains in its ground state, and can be modeled as a single-electron charge qubit with tunnel coupling $\Delta_1$.
Following the derivation in Ref.~\cite{Corrigan.2023.PhysRevApplied.20.064005}, the transmission response for a resonator interacting with the ac-driven charge qubit is given by
\begin{equation}
    \frac{A}{A_0} = \frac{1 - \hbar \, \g / 2 \, \beta \, \tilde{\varepsilon}}{\sqrt{1 + (2 \, \delta\omega_0 / \kappa)^2}},
    \label{eq:Anorm}
\end{equation}
where $\g = - 4 g_c \Delta_1^2 \tilde{\varepsilon} / E_{\text{cq},0}^3$ and $\delta\omega_0 = 8 \hbar g_c^2 \Delta_1^2 / E_{\text{cq},0}^3$ are dynamic longitudinal and dispersive transverse coupling strengths, and $E_{\text{cq},0} = \sqrt{\varepsilon_0^2 + 4 \Delta_1^2}$ is the bare charge qubit excitation energy.
The constant $\beta$ parameterizes direct excitation of the resonator through gate capacitance, and $\kappa$ is the photon loss rate.

To analyze the data in Fig.~\ref{fig:tunnelcoupling}(a), they are fit using Eq.~(\ref{eq:Anorm}), with $\Delta_1$ and $\beta$ as the fitting parameters. 
The value of $\kappa / 2 \pi = 125$~kHz is known from resonance linewidth measurements, and the charge-photon coupling rate, 
\begin{equation}
    \frac{g_c}{2 \pi} = \frac{1}{2 \pi} \frac{\omega_r \, \alpha_\text{C}^\varepsilon}{2 e} \sqrt{\frac{2 Z_r}{h/e^2}} = 3.2~\text{MHz},
    \label{eq:gc}
\end{equation}
is calculated using Eq.~(\ref{eq:alphaC}) and the resonator characteristic impedance, $Z_r \approx 575$~$\Omega$ \cite{Childress.2004.10.1103/physreva.69.042302, Holman.2021.10.1038/s41534-021-00469-0}.
The result is shown by the yellow curve in Fig.~\ref{fig:tunnelcoupling}(a), yielding an estimated tunnel coupling of $\Delta_1 / h = 7.3 \pm 0.6$~GHz for Fig.~\ref{fig:device}(c). 

The data points in Fig.~\ref{fig:tunnelcoupling}(b) show 41 overlaid traces from Fig.~\ref{fig:readout}(d) (light teal) and 25 traces from Fig.~\ref{fig:readout}(e) (dark teal) where $|\varepsilon_\text{pulse}|$ is large enough to generate occupation in $\ket{1}$.
These two datasets have been vertically offset for clarity, and we have shifted traces from the top portion of Fig.~\ref{fig:readout}(e) horizontally to correct for electrostatic device drift during the measurement.
As explained in the main text, two peaks are observed in each dataset, with the additional $A/A_0$ peaks at $\varepsilon_L$ and $\varepsilon_R$ arising from the excited-state resonator couplings $\gL$ and $\gR$.
While these peaks can be fit individually, we find that a straightforward expansion of Eq.~(\ref{eq:Anorm}) to include additional coupling terms enables the pair of peaks in each dataset to be fit together.
In this case,
\begin{equation}
    \frac{A}{A_0} = \frac{1 - \hbar (\g + c_L \, \gL + c_R \, \gR) / 2 \, \beta \, \tilde{\varepsilon}}{\sqrt{1 + (2 \, \delta\omega_0 / \kappa)^2 + (2 \, c_L \, \delta\omega_L / \kappa)^2 + (2 \, c_R \, \delta\omega_R / \kappa)^2}},
    \label{eq:Anormmulticoupling}
\end{equation}
where $\gLR = - 4 g_c \Delta_{3(2)}^2 \tilde{\varepsilon} / E_{\text{cq},L(R)}^3$ and $\delta\omega_{L(R)} = 8 \hbar g_c^2 \Delta_{3(2)}^2 / E_{\text{cq},L(R)}^3$.
The quantity $E_{\text{cq},L(R)} = \sqrt{(\varepsilon_0 - \varepsilon_{L(R)})^2 + 4 \Delta_{3(2)}^2}$ approximates the charge-qubit-like energy splitting of the $\ket{1}$-$\ket{2}$ anticrossing near $\varepsilon_{L(R)}$.
Finally, the parameters $c_L, c_R \in [0,1]$ represent the qubit's time-averaged $\ket{1}$ occupation, which influences the height of the $A/A_0$ peaks measured at $\varepsilon_L$ and $\varepsilon_R$.
Although we have not explicitly characterized these parameters, we find that the simple assumption that $c_L = c_R = 1$ provides a good match to experimental data, corresponding to the qubit's $T_1$ time being long relative to the pulse cycle duration.
It is important to recognize that the strength of each resonator coupling is sharply peaked at a specific anticrossing and suppressed elsewhere.
As a result, in our device---where the QDHQ anticrossings are well separated in $\varepsilon$---a maximum of one coupling channel is active at any given time.

To estimate the QDHQ energy parameters, we fit the data in Fig.~\ref{fig:tunnelcoupling}(b) using Eq.~(\ref{eq:Anormmulticoupling}) with $\varepsilon_0 \rightarrow \varepsilon_m$ and $c_L = c_R =1$, producing the yellow curves.
In fitting the dataset from Fig.~\ref{fig:readout}(d), we fix $\gR = \delta\omega_R = 0$ (since these couplings are inactive), yielding parameter estimates of $\Delta_1 / h = 4.0 \pm 0.2$~GHz, $\Delta_3 / h = 5.4 \pm 0.2$~GHz, and $\varepsilon_L / h = -23.4 \pm 0.4$~GHz.
For the Fig.~\ref{fig:readout}(e) dataset, with $\gL = \delta\omega_L = 0$, we obtain $\Delta_1 / h = 3.7 \pm 0.2$~GHz, $\Delta_2 / h = 6.0 \pm 0.4$~GHz, and $\varepsilon_R = 36.2 \pm 0.6$~GHz.
The two estimates for $\Delta_1$ are in approximate agreement, and we use their average value going forward.

%%%%%%%%%%%%%%%%%%%%%%%%%%%%%%%%%%

\subsection{\texorpdfstring{Right-dot singlet-triplet splitting: $\bm{E_R}$ (pulsed-gate spectroscopy)}{Right-dot singlet-triplet splitting: ER (pulsed-gate spectroscopy)}}

\begin{figure}[t]
    \centering
    \includegraphics[width=0.85\linewidth, trim=0cm 2.5cm 0cm 2.7cm, clip]{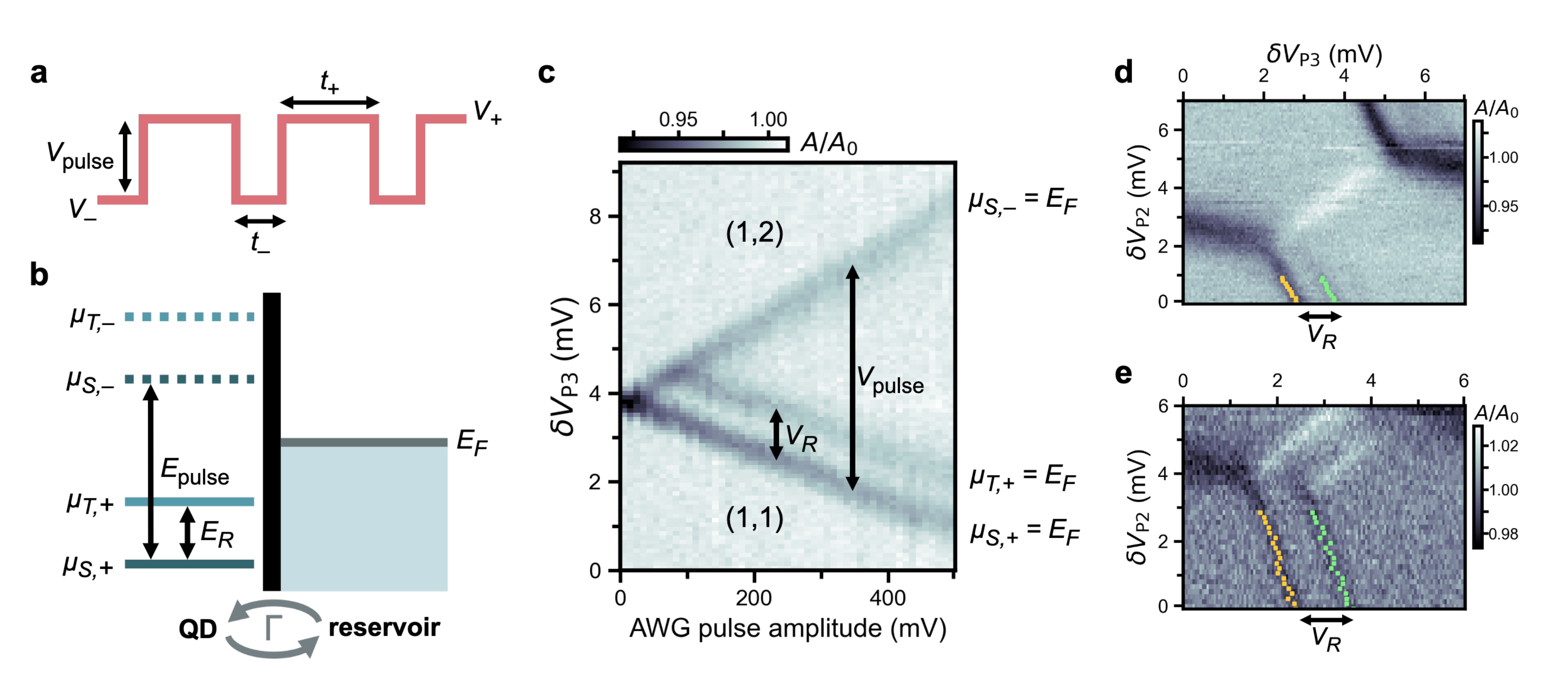}
    \caption{
    Pulsed-gate spectroscopy measurement and analysis.
    \textbf{(a)} Idealized pulsed-voltage waveform applied to gate P3 during measurements.
    \textbf{(b)} Diagram showing cyclic pulsing of the two lowest right-dot energy levels, split by energy $E_R$.
    We label the chemical potential for the spin-singlet ground state $\mu_{S,+}$ during the high-voltage pulse stage and $\mu_{S,-}$ during the low-voltage stage; the spin-triplet, excited-state chemical potential during each stage is  $\mu_{T,+}$ and $\mu_{T,-}$.
    Electrons tunnel cyclically between the dot and neighboring reservoir during pulsing with an overall rate $\Gamma$.
    \textbf{(c)} Typical pulsed-gate spectroscopy measurement: $A/A_0$ as a function of $V_\text{P3}$ and AWG pulse amplitude (measured in instrument units).
    Dips in $A / A_0$ are caused by resonant electron tunneling.
    \textbf{(d)} Data from Fig.~\ref{fig:device}(c) with extracted coordinates for the $\mu_{S,+}$ and $\mu_{T,+}$ tunneling resonances marked with yellow and green points, respectively.
    The voltage splitting $V_R$ between resonances is used to estimate $E_R$.
    \textbf{(e)} Data from Fig.~\ref{fig:readout}(c) with an extended measurement range for the $V_\text{P2}$ axis.
    Tunneling resonance coordinates are marked as in (d) and used to estimate $E_R$.
    }
    \label{fig:pulsespec}
\end{figure}

Pulsed-gate spectroscopy is performed by applying an alternating-voltage pulse train to the plunger P3.
Figure~\ref{fig:pulsespec}(a) depicts an idealized voltage waveform with peak-to-peak amplitude $V_\text{pulse}$ and a duty cycle defined by \hbox{$D \equiv t_+ / (t_+ + t_-)$}, where $t_+$ and $t_-$ are the periods of the high-voltage ($V_+$) and low-voltage ($V_-$) pulse stages. 
The voltage pulses cycle the right-dot energy levels between upper and lower positions separated by $E_\text{pulse} = \alpha_\text{P3}^R V_\text{pulse}$, as diagrammed in Fig.~\ref{fig:pulsespec}(b).

Figure~\ref{fig:pulsespec}(c) shows an illustrative measurement of $A/A_0$ at the (1,1)-(1,2) reservoir transition as a function of $V_\text{P3}$ and arbitrary waveform generator (AWG) pulse amplitude.
We note that this measurement is acquired at a different device tuning than the main-text experiments, and the microwave resonator drive is delivered through an alternative on-chip port rather than through P3.
The duty cycle of the applied pulse train is $D = 0.6$, with $t_+ = 3$~ns and $t_- = 2$~ns.
The reservoir energy barrier and pulse-train frequency are calibrated so that the total electron tunneling rate $\Gamma$ sharply increases at tunings where a dot energy level aligns with the reservoir Fermi energy $E_F$.
The onset of resonant tunneling at these points changes the dot's complex admittance, thereby suppressing $A / A_0$.
The features in Fig.~\ref{fig:pulsespec}(c) closely resemble those of time-averaged pulsed-gate spectroscopy measurements using conventional integrated charge sensors and lock-in techniques (e.g., Ref~\cite{Elzerman.2004.10.1063/1.1757023}).
For AWG pulse amplitudes below $\sim$100 mV, two tunneling resonances are visible, corresponding to P3 voltages where $\mu_{S,+} = E_F$ and $\mu_{S,-} = E_F$.
The splitting between these features provides a direct measurement of the on-chip  amplitude $V_\text{pulse}$. 
At larger amplitudes, the pulse window straddles the first excited energy level, and a third resonance, corresponding to $\mu_{T,+} = E_F$, is observed.
The singlet-triplet splitting,
\begin{equation}
    E_R = \mu_{T,+} - \mu_{S,+} = \alpha_\text{P3}^R V_R,
    \label{eq:ERpulsespec}
\end{equation}
may be estimated from this measurement using the voltage splitting $V_R$ between the $\mu_{S,+}$ and $\mu_{T,+}$ tunneling resonances.

The pulse train applied in Fig.~\ref{fig:device}(c) of the main text is configured with $V_\text{pulse} \approx 1.7$~mV, $t_+ = 15$~ns, and $t_- = 2$~ns.
(As an aside, we speculate that no $A / A_0$ boost from the $\gR$ resonator coupling is observed in this experiment because the duty cycle $D = 0.88$ is too small to provide sufficient signal integration time at $\varepsilon_R$.)
To estimate $E_R$, voltage coordinates of the tunneling resonances are extracted using phenomenological double-Lorentzian fits.
These coordinates are overlaid on the measurement data in Fig.~\ref{fig:pulsespec}(d), with yellow pixels corresponding to the $\mu_{S,+}$ resonance and green pixels corresponding to the $\mu_{T,+}$ resonance.
The $\mu_{S,-}$ resonance is faintly visible as well. 
We estimate $V_R = 0.97 \pm 0.03$~mV from the average voltage splitting between resonance coordinates, which equates to $E_R = 34.8 \pm 1.2$~GHz using Eq.~(\ref{eq:ERpulsespec}).

Reservoir tunneling resonances are also observed in the lower part of Fig.~\ref{fig:readout}(c).
The control pulse waveform used for this readout experiment has a very large duty cycle of $D = 0.96$, with $t_+ = t_\text{meas} = 48$~ns and $t_- = t_\text{pulse} = 2$~ns.
Only the resonances associated with $\mu_{S,+}$ and $\mu_{T,+}$ tunneling are visible due to the extreme duty cycle.
Figure~\ref{fig:pulsespec}(e) shows the same measurement as Fig.~\ref{fig:readout}(c), but with an extended $V_\text{P2}$ sweep range.
We extract $V_R = 1.15 \pm 0.08$~mV and $E_R = 41.3 \pm 2.9$~GHz from the tunneling resonances using the analysis method described above.
This $E_R$ value is larger than the value measured in Fig.~\ref{fig:pulsespec}(d).
The discrepancy is likely due to variation in electrostatic device tuning between the measurements; singlet-triplet splitting has been shown to change substantially as a function of quantum-dot confinement and lateral position due to local interface disorder \cite{Dodson.2022.10.1103/physrevlett.128.146802}.

Finally, we note that we are only able to implement pulsed-gate spectroscopy for the right quantum dot in this device, as the left-dot plunger P2 is not connected to coaxial wiring for high-frequency pulsing.
The left dot's singlet-triplet splitting $E_L$ is therefore estimated based on other measurement data and numerical calculations described in the next section.

%%%%%%%%%%%%%%%%%%%%%%%%%%%%%%%%%%

\subsection{\texorpdfstring{Other Hamiltonian parameters: $\bm{E_L, \, \Delta_4}$}{Other Hamiltonian parameters: EL, Δ4}}

\begin{figure}[t]
    \centering
    \includegraphics[width=1.0\linewidth]{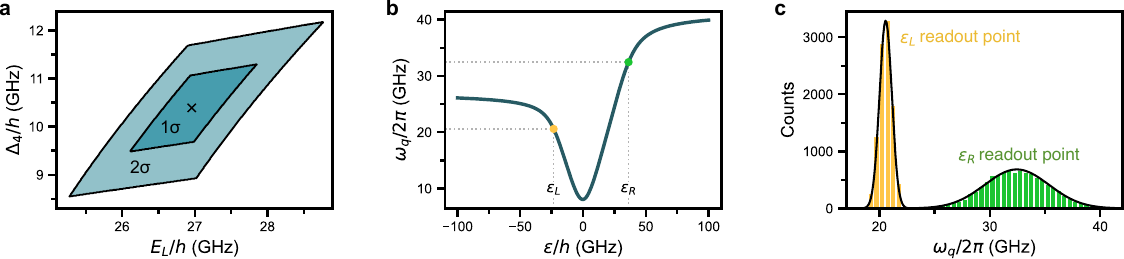}
    \caption{
    Qubit spectroscopy numerical analysis.
    \textbf{(a)} Contours showing regions of $(E_L, \Delta_4)$ parameter space in which calculated $\varepsilon_L$ and $\varepsilon_R$ values match Fig.~\ref{fig:readout} measurements to within $1\sigma$ or $2\sigma$ experimental uncertainty.
    The closest-match parameter coordinates marked with an `$\times$' are used as $E_L$ and $\Delta_4$ estimates for main-text experiments.
    \textbf{(b)} Qubit frequency $\omega_q$ as a function of $\varepsilon$.
    Yellow and green circles mark frequencies at the $\varepsilon_L$ and $\varepsilon_R$ readout operating points.
    \textbf{(c)} Results of a Monte Carlo simulation calculating qubit frequency at the $\varepsilon_L$ (yellow histogram) and $\varepsilon_R$ (green histogram) readout operating points. 
    Black curves show Gaussian fits to the histograms.
    }
    \label{fig:numerics}
\end{figure}

The main-text experiments do not provide direct measurements of the left-dot singlet-triplet splitting $E_L$ or the tunnel coupling $\Delta_4$. 
However, we are able to estimate these parameters through numerical analysis.
We first diagonalize the qubit Hamiltonian in Eq.~(\ref{eq:H_qdhq}) as a function of $E_L$ and $\Delta_4$ with the other matrix entries fixed to the values derived above. 
Then, we compute the $\ket{1}$-$\ket{2}$ energy anticrossing detunings $\varepsilon_L$ and $\varepsilon_R$ for each diagonalized matrix and compare them to the experimentally measured values.
Figure~\ref{fig:numerics}(a) summarizes the results of this analysis, with color-filled contours bounding the regions of $(E_L, \Delta_4)$ parameter space in which the calculated $\varepsilon_L$ and $\varepsilon_R$ values match experimental results to within $1\sigma$ or $2\sigma$ experimental uncertainty.
The closest match to our measurements occurs at the coordinate pair marked with an `$\times$' in the plot; from these coordinates, we estimate the qubit parameters in the Fig.~\ref{fig:readout} measurements to be $E_L / h = 27.0 \pm 0.4$~GHz and $\Delta_4 / h = 10.4 \pm 0.7$~GHz (the uncertainties are given by the $1\sigma$ contour dimensions about this point).

%%%%%%%%%%%%%%%%%%%%%%%%%%%%%%%%%%

\subsection{Qubit frequency at the readout operating points}

With all QDHQ parameters estimated, the qubit frequency $\omega_q$ can be calculated by diagonalizing the Hamiltonian in Eq.~(\ref{eq:H_qdhq}).
Figure~\ref{fig:numerics}(b) plots $\omega_q$ as a function of $\varepsilon$.
For our work, we are particularly interested in the qubit frequency at the readout operating points marked by green and yellow circles in the plot.
Using the Hamiltonian parameters for the Fig.~\ref{fig:readout} measurements found above, we compute $\omega_q / 2 \pi = 20.6 \pm 0.6$~GHz at the $\varepsilon_L$ readout point and $\omega_q / 2 \pi = 32.4 \pm 2.9$~GHz at the $\varepsilon_R$ readout point.

The qubit frequency uncertainties are obtained through a Monte Carlo simulation.
In each simulation iteration, Hamiltonian parameter entries are drawn from Gaussian distributions whose means and standard deviations correspond to the values and uncertainties computed in previous sections.
Figure~\ref{fig:numerics}(c) charts histograms of the readout-point qubit frequencies after 10,000 iterations.
The frequency uncertainties are extracted through Gaussian fits to the histograms.

%%%%%%%%%%%%%%%%%%%%%%%%%%%%%%%%%%
% COUPLING STRENGTHS
%%%%%%%%%%%%%%%%%%%%%%%%%%%%%%%%%%

\section{Qubit-Resonator Coupling Strengths \& Measurement Rate}

To provide a representative estimate of the longitudinal and transverse interaction strengths in this work, we adopt an approach similar to that in Ref.~\cite{Corrigan.2023.PhysRevApplied.20.064005}, using parameter values obtained by fitting the data in Fig.~\ref{fig:tunnelcoupling}(a).

The transverse coupling $\delta\omega_0$ is heavily suppressed in the ultra-dispersive regime.
At $\varepsilon_0 = 0$, where the transverse interaction is maximized, we calculate a coupling strength of only
\begin{equation}
    \frac{\delta\omega_0}{2\pi} = \frac{1}{2\pi} \frac{\hbar g_c^2}{\Delta_1} = 1.4~\text{kHz}.
    \label{eq:dispersive_coupling_strength}
\end{equation}

The longitudinal coupling $\g$ depends on the ac drive applied to the qubit.
Figure~\ref{fig:tunnelcoupling}(a) is measured using a microwave source output power of 9 dBm.
The 50~$\Omega$ coaxial cabling on the drive line is configured with 39~dB of attenuation at room temperature and 40~dB within the cryostat.
Based on SPICE simulations, we estimate an additional 10~dB of attenuation due to impedance mismatch at the device gate leads, which are buried coplanar waveguides engineered to have a low characteristic impedance of $Z_g \approx 1~\Omega$ \cite{Holman.2021.10.1038/s41534-021-00469-0, Corrigan.2023.PhysRevApplied.20.064005}.
Factoring in this attenuation, the on-chip drive power is estimated to be $P = -80~\text{dBm} = 10$~pW, which generates an ac detuning amplitude of $\tilde{\varepsilon} / h = \alpha_\text{P3}^\varepsilon \sqrt{2 Z_g P} / h = 113.3$~MHz.
Using this value, we compute 
\begin{equation}
    \frac{|\g|}{2\pi} = \frac{1}{2\pi} \frac{g_c \tilde{\varepsilon}}{2 \Delta_1} = 24.6~\text{kHz}
    \label{eq:longitudinal_coupling_strength}
\end{equation}
at $\varepsilon_0 = 0$, where the coupling is at its strongest.

Based on these values, we expect that longitudinal coupling dominates the quantum measurement rate $\Gamma_\text{meas}$ in our experiments, such that $\Gamma_\text{meas} / 2 \pi \approx \tilde{g}_\parallel^2 / 2 \pi \kappa \sim 5$~kHz \cite{Ruskov.2019.10.1103/physrevb.99.245306, ruskov.2024.longitudinal}.
We note that $\Gamma_\text{meas}$ can be substantially increased by enhancing $g_c$ and $\tilde{\varepsilon}$, while simultaneously optimizing $\kappa$ and the relevant tunnel couplings. 
For example, the charge-photon coupling $g_c$ can be engineered to be more than 10 times larger (as demonstrated experimentally in, e.g., Refs.~\cite{Mi.2018.coherent, Dijkema.2023.Two-qubit}), while maintaining $\tilde{\varepsilon}/\hbar g_c \gg 1$.

We emphasize that the coupling values above should be treated as rough estimates, given that they are not directly measured, and the microwave environment of the device is not fully characterized.
The key takeaway is that the longitudinal coupling is found to be an order of magnitude larger than the residual transverse coupling.
Based on the $A/A_0$ peaks observed in our measurements and the parameter values reported in Table~\ref{tab:params}, we expect that longitudinal coupling serves as the primary qubit-resonator interaction mechanism for all experiments.

%%%%%%%%%%%%%%%%%%%%%%%%%%%%%%%%%%
% PARAMETER TABLE
%%%%%%%%%%%%%%%%%%%%%%%%%%%%%%%%%%

\section{Parameter Table}

\begin{table}[H]
    \renewcommand{\arraystretch}{1.2}
    \centering
    \caption{
    Summary of device and measurement parameters used throughout this work.
    Parameters are classified into one of three categories: `controlled' variables with values chosen as part of the experiment protocol; `measured' values determined primarily through direct measurement; and `calculated' values determined primarily through a numerical calculation or simulation described in previous sections. 
    }
    \begin{tabular}{l@{\hspace{10mm}}c@{\hspace{10mm}}r@{\hspace{3mm}}l}
        \hline\hline
        \textbf{Resonator properties} & \textbf{Determination} & \textbf{Value} & \textbf{Units} \\
        \hline
        Fundamental-mode frequency, $\omega_r / 2 \pi $ & measured & $1.30380$ & GHz \\
        Photon decay rate, $\kappa / 2 \pi$ & measured & $124.5 \pm 0.5$ & kHz \\
        Loaded quality factor, $Q_L$ & measured & $10,476 \pm 46$ & \\
        Characteristic impedance, $Z_r$ & calculated & 575 & $\Omega$ \\
        Charge-photon coupling rate, $g_c / 2 \pi$ & calculated & $3.16 \pm 0.55$ & MHz \\
        \hline
        \textbf{DQD properties} \\
        \hline
        P2-left dot lever arm, $\alpha_\text{P2}^L / h$ & measured & $33.9 \pm 1.2$ & GHz/mV \\
        P3-right dot lever arm, $\alpha_\text{P3}^R / h$ & measured & $36.0 \pm 0.7$ & GHz/mV \\
        P3 detuning lever arm, $\alpha_\text{P3}^\varepsilon / h$ & measured & $25.4 \pm 1.2$ & GHz/mV \\
        Coupling-gate detuning lever arm, $\alpha_\text{C}^\varepsilon / h$ & measured & $5.6 \pm 1.0$ & GHz/mV \\
        Electron temperature, $T_e$ & measured & $213 \pm 6$ & mK \\
        \hline
        \textbf{Fig.~\ref{fig:device}(c)} \\
        \hline
        Microwave source output power & controlled & 9 & dBm \\
        Pulse peak-to-peak voltage amplitude, $V_\text{pulse}$ & controlled & 1.7 & mV \\ 
        Pulse-sequence high-voltage period, $t_+$ & controlled & 15 & ns \\
        Pulse-sequence low-voltage period, $t_-$ & controlled & 2 & ns \\
        Pulse-sequence duty cycle, $D$ & controlled & 0.88 \\
        Pulse-sequence repetitions per data point & controlled & $8.8 \times 10^6$ \\
        Interdot tunnel coupling, $\Delta_1 / h$ & measured & $7.3 \pm 0.6$ & GHz \\
        Right-dot tunneling resonance voltage splitting, $V_R$ & measured & $0.97 \pm 0.03$ & mV \\
        Right-dot singlet-triplet splitting, $E_R / h$ & measured & $34.8 \pm 1.2$ & GHz \\
        Minimum qubit frequency, $\omega_q(\varepsilon_0 = 0) / 2 \pi$ & calculated & $14.6 \pm 1.2$ & GHz \\
        Maximum qubit frequency, $\omega_q(\varepsilon_0 \gg 0) / 2 \pi$ & calculated & $34.8 \pm 1.2$ & GHz \\
        Transverse coupling strength, $\delta\omega_0(\varepsilon_0 = 0)/2\pi$ & calculated & 1.4 & kHz \\
        Longitudinal coupling strength, $|\g(\varepsilon_0 = 0)|/2\pi$ & calculated & 24.6 & kHz \\
        \hline\
        \textbf{Figs.~\ref{fig:readout}(b, c, d, e)} \\
        \hline
        Microwave source output power (b, c) & controlled & 5 & dBm \\
        Microwave source output power (d, e) & controlled & 7 & dBm \\
        Pulse detuning amplitude (b, c), $|\varepsilon_\text{pulse}| / h$ & controlled & 63.5 & GHz \\
        Pulse detuning amplitude (d, e), $|\varepsilon_\text{pulse}| / h$ & controlled & 0 to 63.5 & GHz \\
        Pulse phase period, $t_\text{pulse}$ ($= t_-$) & controlled & 2 & ns \\
        Measurement phase period, $t_\text{meas}$ ($= t_+$) & controlled & 48 & ns \\
        Pulse-sequence duty cycle, $D$ & controlled & 0.96 \\ 
        \hline\hline
    \end{tabular}
    \label{tab:params}
\end{table}

\begin{table}[H]
    \renewcommand{\arraystretch}{1.2}
    \centering
    \begin{tabular}{l@{\hspace{10mm}}c@{\hspace{10mm}}r@{\hspace{3mm}}l}
        \hline\hline
        \textbf{Figs.~\ref{fig:readout}(b, c, d, e)} & \textbf{Determination} & \textbf{Value} & \textbf{Units} \\
        \hline
        Pulse-sequence repetitions per data point (b, c) & controlled & $4 \times 10^6$ \\
        Pulse-sequence repetitions per data point (d, e) & controlled & $2 \times 10^6$ \\
        Interdot tunnel coupling, $\Delta_1 / h$ & measured & $3.9 \pm 0.1$ & GHz \\
        Interdot tunnel coupling, $\Delta_2 / h$ & measured & $6.0 \pm 0.4$ & GHz \\
        Interdot tunnel coupling, $\Delta_3 / h$ & measured & $5.4 \pm 0.2$ & GHz \\
        Interdot tunnel coupling, $\Delta_4 / h$ & calculated & $10.4 \pm 0.7$ & GHz \\
        Left-dot singlet-triplet splitting, $E_L / h$ & calculated & $27.0 \pm 0.4$ & GHz \\
        Right-dot tunneling resonance voltage splitting, $V_R$ & measured & $1.15 \pm 0.08$ & mV \\
        Right-dot singlet-triplet splitting, $E_R / h$ & measured & $41.3 \pm 2.9$ & GHz \\
        $\ket{1}$-$\ket{2}$ anticrossing detuning, $\varepsilon_L / h$ & measured & $-23.4 \pm 0.4$ & GHz \\
        $\ket{1}$-$\ket{2}$ anticrossing detuning, $\varepsilon_R / h$ & measured & $36.2 \pm 0.6$ & GHz \\
        Minimum qubit frequency, $\omega_q(\varepsilon_0 = 0) / 2 \pi$ & calculated & $7.8 \pm 0.2$ & GHz \\
        Maximum qubit frequency, $\omega_q(\varepsilon_0 \gg 0) / 2 \pi$ & calculated 
        & $41.3 \pm 2.9$ & GHz \\
        Readout-point qubit frequency, $\omega_q(\varepsilon_0 = \varepsilon_L) / 2 \pi$ & calculated & $20.6 \pm 0.6$ & GHz \\
        Readout-point qubit frequency, $\omega_q(\varepsilon_0 = \varepsilon_R) / 2 \pi$ & calculated & $32.4 \pm 2.9$ & GHz \\
        \hline\hline
    \end{tabular}
\end{table}

\clearpage

%%%%%%%%%%%%%%%%%%%%%%%%%%%%%%%%%%
% EXPERIMENTAL SETUP
%%%%%%%%%%%%%%%%%%%%%%%%%%%%%%%%%%

\clearpage

\section{Experimental Setup}

\begin{figure}[h]
    \centering
    \includegraphics[width=0.94\linewidth, trim=0cm 1.5cm 0cm 4.7cm, clip]{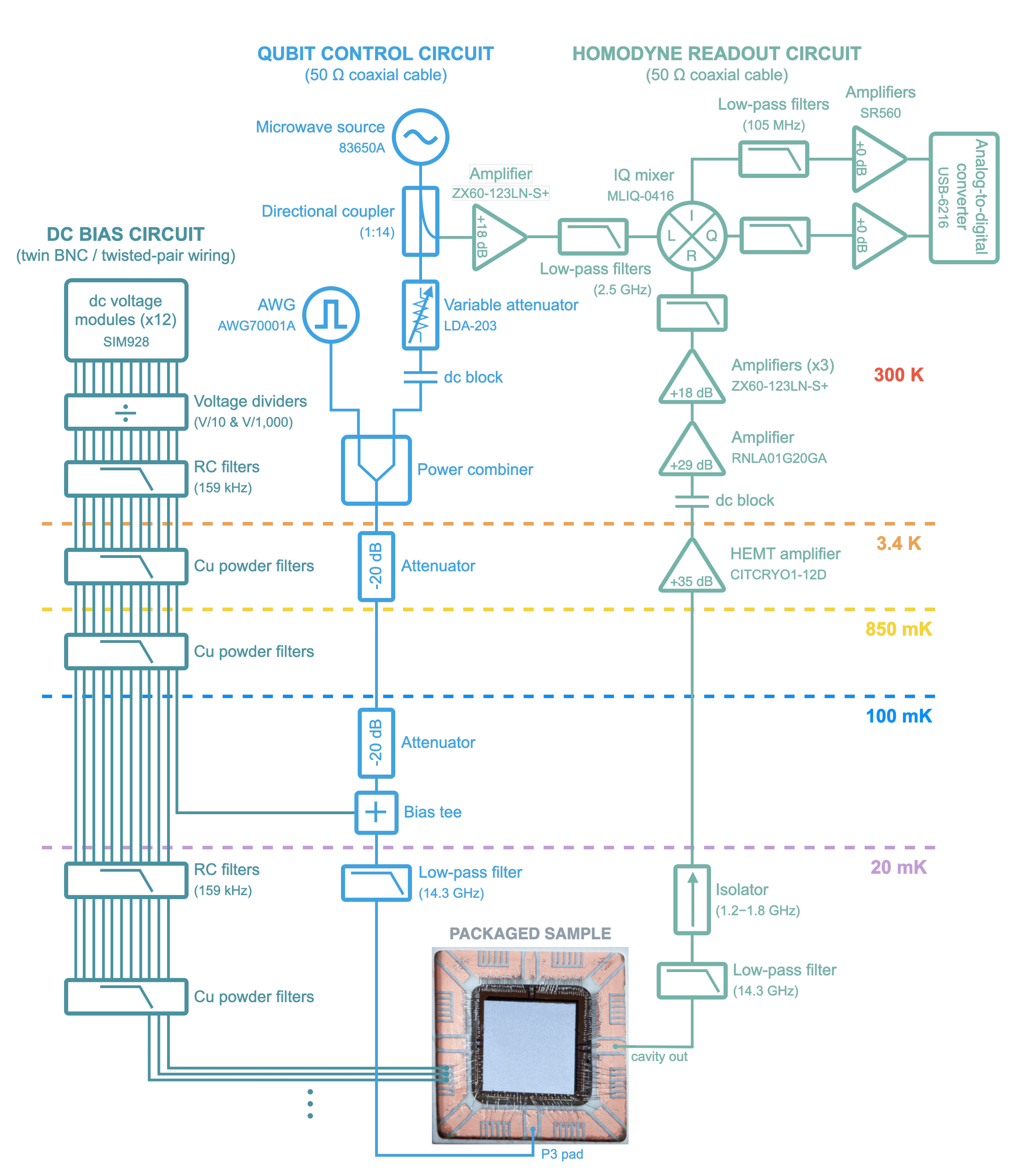}
    \caption{
    Wiring diagram for the measurement instrumentation. 
    Experiments are performed in a Leiden Cryogenics CF-450 dilution refrigerator with a mixing-chamber base temperature around 20 mK.
    }
    \label{fig:wiring}
\end{figure}

\end{document}